\documentclass[%
 reprint,
 amsmath,amssymb,
 aps,
]{revtex4-2}

\usepackage{graphicx}
\usepackage{dcolumn}
\usepackage{bm}
\usepackage{hyperref}
\usepackage[export]{adjustbox}
\usepackage{hyperref}

\begin{document}

\preprint{APS/123-QED}

\title{ Development of Low-Threshold Detectors for Low-Mass Dark Matter Searches Using an N-Type Germanium Detector at 5.2 K }

\author{Sanjay Bhattarai}
 \altaffiliation[Also at ]{Physics Department, University of South Dakota.}
\author{Dongming Mei}%
 \email{dongming.mei@usd.edu}
 \author{Rajendra Panth}
 \author{Mathbar Singh Raut}
 \author{Kyler Kooi}
 \author{Hao Mei}
 \author{Guojian Wang}
\affiliation{%
 University of South Dakota, Vermillion, SD, 57069\\
}%


\date{\today}

\begin{abstract}
We investigated charge transport in an n-type germanium detector at 5.2 K to explore new technology for enhancing low-mass dark matter detection sensitivity. Calculations of dipole and cluster dipole state binding energies and electric field-dependent trapping cross-sections are critical to developing low-threshold detectors. The detector operates in two modes: depleting at 77K before cooling, or directly cooling to 5.2 K and applying different bias voltages. Results indicated lower binding energy of charge states in the second mode, at zero field and under an electric field, suggesting different charge states formed under different operating modes. Measured cluster dipole and dipole state binding energies at zero field were 7.884±0.644 meV and 8.369±0.748 meV, respectively, signifying high low-threshold potential for low-mass dark matter searches in the future.

\end{abstract}

\maketitle


    \section{\label{sec:level1}Introduction}

The interaction between dark matter (DM) and ordinary matter is limited to weak elastic scattering processes, resulting in only a small energy deposition from nuclear or electron recoil~\cite{ahmed2011results,armengaud2012search,zhao2013first}. This highlights the need for a detector with a very low energy threshold to detect DM~\cite{mei2018direct}. The LZ experiment has pushed the sensitivity for weakly interacting massive particles (WIMPs) with a mass greater than 10 GeV/c$^2$ to the point where the neutrino-induced background limits its sensitivity~\cite{aalbers2022first}. However, the recent emergence of low-mass DM in the MeV range has generated excitement as a DM candidate, although current experiments cannot detect it due to its small mass. The detection of MeV-scale DM requires new detectors with thresholds as low as sub-eV, since both electronic and nuclear recoils from MeV-scale DM range from sub-eV to 100 eV~\cite{essig2012direct}. Conventional detector techniques cannot detect this low-mass DM.

Germanium (Ge) detectors have the lowest energy threshold among any current detector technology, making them ideal for low-mass DM searches~\cite{agnese2014search,agnese2019search,armengaud2012search,armengaud2018searches}. The band gap of Ge at 77K is 0.7 eV and the average energy required to generate an electron-hole pair in Ge is about 3 eV~\cite{wei2017average}. Thus, a Ge detector can provide a very low energy threshold. Furthermore, proper doping of the Ge detector with impurities can expand the parameter space for low-mass DM searches even further. Shallow-level impurities in Ge detectors have binding energies of about 0.01 eV, and can form dipole states and cluster dipole states when operated at temperatures below 10 K~\cite{mei2018direct,bhattarai2021low,mei2022evidence}. These dipole states and cluster dipole states have even lower binding energies than the impurities themselves, providing a potential avenue for detecting low-mass DM. Although the binding energies of impurities in Ge is well understood~\cite{venos2000behaviour,sundqvist2009measurement}, little is known about the binding energy of the dipole states and cluster dipole states near helium temperature.

At low temperatures near liquid helium, residual impurities in germanium freeze out from the conduction or valence band into localized states, forming electric dipoles ($D^{0^{*}}$ for donors and $A^{0^{*}}$ for acceptors) or neutral states ($D^0$ and $A^0$). These dipole states have the ability to trap charge carriers and form cluster dipole states ($D^{+^{*}}$ and $D^{-^{*}}$ for donors, and $A^{+^{*}}$ and $A^{-^{*}}$ for acceptors)\cite{mei2022evidence}. This phenomenon has been studied in detail in a previous work by Mei et. al\cite{mei2022evidence}. When an alpha particle ($\alpha$) from an $^{241}$Am decay is sent to a Ge detector, it deposits energy and creates electron-hole pairs within a 10 $\mu$m range from the surface of the detector~\cite{ziegler2010srim,arnquist2022alpha}. By applying a positive or negative bias voltage to the bottom of the detector and operating it at a cryogenic temperature of approximately 4 K, only one type of charge carrier is drifted through the detector. These drifted charge carriers undergo a dynamic process of elastic scattering, trapping, and de-trapping, allowing us to study the binding energy of the formed dipole states and cluster dipole states. In this study, an n-type Ge detector is operated in two different modes, applying different bias voltages and cooling the detector to cryogenic temperature.

\subsection{Mode 1}

In this mode, an n-type planar detector is first cooled to 77K and a bias voltage is applied, gradually increasing until the detector is fully depleted. The bias is then increased by an additional 600 volts to become the operational voltage. The detector is then cooled down to 5.2 K while still under the applied operational voltage. At 77 K, the depletion process causes all the free charge carriers to be swept away, leaving only the space charge states, $D^+$, behind. Upon cooling to 5.2 K, a charge trapping process occurs, resulting in the formation of dipole states as electrons drift across the detector~\cite{mei2022evidence}. Continued drift of electrons across the detector can result in de-trapping of charge carrier through impact ionization of the dipole states. The key charge-trapping and de-trapping processes are described below:
\begin{equation}
    e^- + D^{+} \rightarrow D^{0^{*}}, 
    e^-+D^{0^{*}} \rightarrow 2e^- +D^+
    \label{eq:my_label2}.
    \end{equation}

In this mode, the operation of the n-type planar detector begins with the formation of dipole states via charge trapping as a result of the Coulomb force between the space charge states and the drifting electrons. The second process is the release of trapped charge through impact ionization of the dipole states, known as charge de-trapping. By examining the time-dependent behavior of this de-trapping process, we are able to determine the binding energy of the dipole states.

\subsection{Mode 2}

In this mode of operation, the n-type planar Ge detector is cooled directly to 5.2 K without any applied bias voltage. Once cooled, the detector is then biased to the desired voltage level. At these low temperatures, impurities in the Ge crystal freeze out from the conduction or valence band to form localized states that result in the creation of dipole states. As it is an n-type detector, the majority of these dipole states are $D^{0^{*}}$~\cite{mei2022evidence}. When an $\alpha$ source is placed near the detector, the resulting $\alpha$-particle-induced electron-hole pairs are created on the surface of the detector. Upon applying a positive bias voltage to the bottom of the detector, the electrons created by the $\alpha$ particles are drifted across the detector, leading to the following processes occurring within the detector:

\begin{equation}
  e^-+D^{0^{*}}\rightarrow D^{-^{*}}, 
  e^-+D^{-^{*}}\rightarrow 2e^- + D^{0^{*}}
  \label{eq:my_label1}.
    \end{equation}

The first process in this mode is a trapping of charges by the Coulomb forces exerted by the dipole states on the drifted electrons, resulting in the formation of cluster dipole states. The second process is a de-trapping of charges through impact ionization of the cluster dipole states. The detector experiences a dynamic process of charge trapping, transport, and creation. The study of the time-dependent de-trapping of charges through the impact ionization of cluster dipole states helps us determine their binding energy.

When comparing the two operational modes, it can be noted that in Mode 2, the dipole states are formed at 5.2 K without any applied bias voltage. These dipole states rapidly trap charges as soon as the electrons are drifted across the detector, resulting in a shorter trapping time and lower binding energy. In contrast, in Mode 1, the dipole states are formed in the space charge region when electrons are drifted across the detector with an applied bias voltage. Therefore, it is expected that the trapping time will be longer and the binding energy of the dipole states will be higher than that of the cluster dipoles.

\subsection{Physics Model}

As mentioned earlier, the formation of dipole states and cluster dipole states in the detector depends on the operational mode. In Mode 2, when the n-type Ge detector is cooled down to 5.2 K, the majority impurity atoms freeze out from the conduction band and form electric dipole states, $D^{0^{*}}$. If a positive bias voltage is applied to the bottom of the detector, electrons produced by the $\alpha$ particles from the $^{241}$Am source, which is located above the detector within the cryostat, can be drifted across the detector. This drifting of electrons leads to the formation of cluster dipole states, D$^{-^{*}}$, through the charge trapping between the dipole states and the drifted electrons. As the bias voltage increases, the charge carriers gain more kinetic energy and begin to emit from the traps, resulting in a decrease in the number of cluster dipole states and an increase in electric dipole states.

In Mode 1, when a positive bias voltage is applied, electrons are drifted across the detector, leading to the formation of dipole states D$^{0^{*}}$ through the space charge states of D$^{+}$. As the bias voltage increases, the drifted electrons gain more kinetic energy and are capable of freeing trapped electrons from the dipole states. In both modes, the emission rate of the charge carriers is time-dependent and reaches a balance when the charge emission and charge trapping are equal. At a sufficient bias voltage, such as around 800 volts, charge trapping becomes negligible and the charge emission also becomes negligible. The emission rate ($e_{n}$) of the charge carriers can be mathematically expressed as:~\cite{lee1999compensation}.
\begin{equation}
e_n=\sigma_{trap} v_{th} N_c \exp \left(-\frac{E_B}{k_BT}\right),
\label{eq:my_label3}
\end{equation}

where $\sigma_{trap}$ represents the trapping cross-section, $v_{th}$ is the thermal velocity, $N_c$ = 2.46 $\times$ 10$^{15}$/cm$^3$ is the effective density of states of electrons in the conduction band at 5.2 K, $E_B$ is the binding energy of the trapped charge carriers, $k_B$ is the Boltzmann constant, and $T$ is the temperature of the detector.

By using the experimental data to directly determine $e_{n}$ and by knowing the values of $v_{th}$, $N_c$, and $T$, one can obtain the binding energy of dipole states or cluster dipole states from equation~\ref{eq:my_label3}, provided the value of the trapping cross-section, $\sigma_{trap}$, is known. However, determining the value of $\sigma_{trap}$ requires further calculation, as will be discussed.

The trapping cross-section ($\sigma_{trap}$) of the charge carriers is related to the trapping length ($\lambda_{th}$) through the following relation:\cite{phipps2016ionization, mei2020impact}
\begin{equation}
\lambda_{th}=\frac{1}{\left(\frac{N_A + N_D \pm \left| N_A - N_D \right|}{2}\right)\times \left(\sigma_{trap}\times\frac{v_{tot}}{v_d}\right)},
\label{eq:my_label4}
\end{equation}
where $N_{A}$ and $N_{D}$ represent the p-type and n-type impurities, respectively. $v_{tot}$ is the total velocity of the drift electrons, and $v_d$ is the drift velocity, which is dependent on the electric field ($E$) and is given by:
\begin{equation}
v_d\approx \frac{\mu_0E}{1+\mu_0 E/v_{sat}},
\label{eq:my_label5}
\end{equation}
where $\mu_0$ represents the mobility of the charge carrier when the field is zero, and can be expressed as $\mu_0=\mu_0(H)/r$. The Hall mobility, $\mu_0(H)$, has standard values of $36000\text{ cm}^2/\text{Vs}$ for electrons and $42000\text{ cm}^2/\text{Vs}$ for holes, while the corresponding values of $r$ are 0.83 for electrons and 1.03 for holes. The saturation velocity, $v_{sat}$, can be calculated using the following empirical formula\cite{mei2020impact}:
\begin{equation}
v_{sat} = \frac{v_{sat}^{300}}{1-A_v + A_v(T/300)}.
\label{eq:my_label6}
\end{equation}
The saturation velocity at 300 K, $v_{sat}^{300}$, for electrons and holes are $7\times10^6\text{ cm/s}$ and $6.3\times 10^6\text{ cm/s}$, respectively. The values of $A_v$ for electrons and holes are 0.55 and 0.61, respectively~\cite{quay2000temperature}.
Additionally, the charge collection efficiency ($\epsilon$) of a planar Ge detector can be related to the trapping length ($\lambda_{th}$) through the following formula~\cite{he2001review,mei2020impact}:
\begin{equation}
\epsilon = \frac{\lambda_{th}}{L}\left(1 - \text{exp}\left(-\frac{L}{\lambda_{th}}\right)\right),
\label{eq:my_label7}
\end{equation}
where $L$ = 5.5 mm represents the detector thickness.

The determination of the charge collection efficiency ($\epsilon$) in a planar Ge detector enables us to calculate the charge trapping cross-section ($\sigma_{trap}$) using equation~\ref{eq:my_label4}. The necessary inputs, such as the net impurity concentration ($N_A + N_D \pm |N_A-N_D|$), are known from the Hall effect and capacitance-voltage measurements, while the electric field ($E$) in the detector can be obtained using the applied bias voltage.

With the calculated values of $\epsilon$ and the known thickness of the detector (L), we can find $\lambda_{th}$ from equation~\ref{eq:my_label7}. The total velocity ($v_{tot}$) of the charge carriers is the combination of their thermal velocity ($v_{th}$) and the saturation velocity ($v_{sat}$). By combining the equations for $\lambda_{th}$ and $v_{tot}$, we can determine the electric field-dependent trapping cross-section ($\sigma_{trap}$)~\cite{mei2020impact}.

In an n-type Ge detector, the emission rate ($e_n$) of charge carriers from the traps is measured during operation in both Mode 1 and Mode 2. The energy versus time plot is used to determine the emission rate by analyzing the slope of the plot after a given bias voltage has been applied to the detector. By combining this value with equation~\ref{eq:my_label1}, we can find the binding energy of dipole states and cluster dipole states in the n-type Ge detector at cryogenic temperature.

\section{\label{sec:level2}Experimental procedure}
The USD crystal growth and detector development infrastructure is a state-of-the-art facility equipped with a zone refining process for purifying commercial ingots to a high level of purity suitable for crystal growth using the Czochralski method~\cite{yang2014investigation,wang2012development,raut2020characterization}. This results in high-quality homegrown crystals that are used for the fabrication of n-type (R09-02) detectors in the USD detector fabrication lab~\cite{panth2022temperature}. The R09-02 detector has a net impurity concentration of $7.02\times 10^{10}/cm^3$ and dimensions of 11.7 mm $\times$ 11.5 mm $\times$ 5.5 mm.

\begin{figure}
    \centering
    \includegraphics[width=0.55\textwidth,inner]{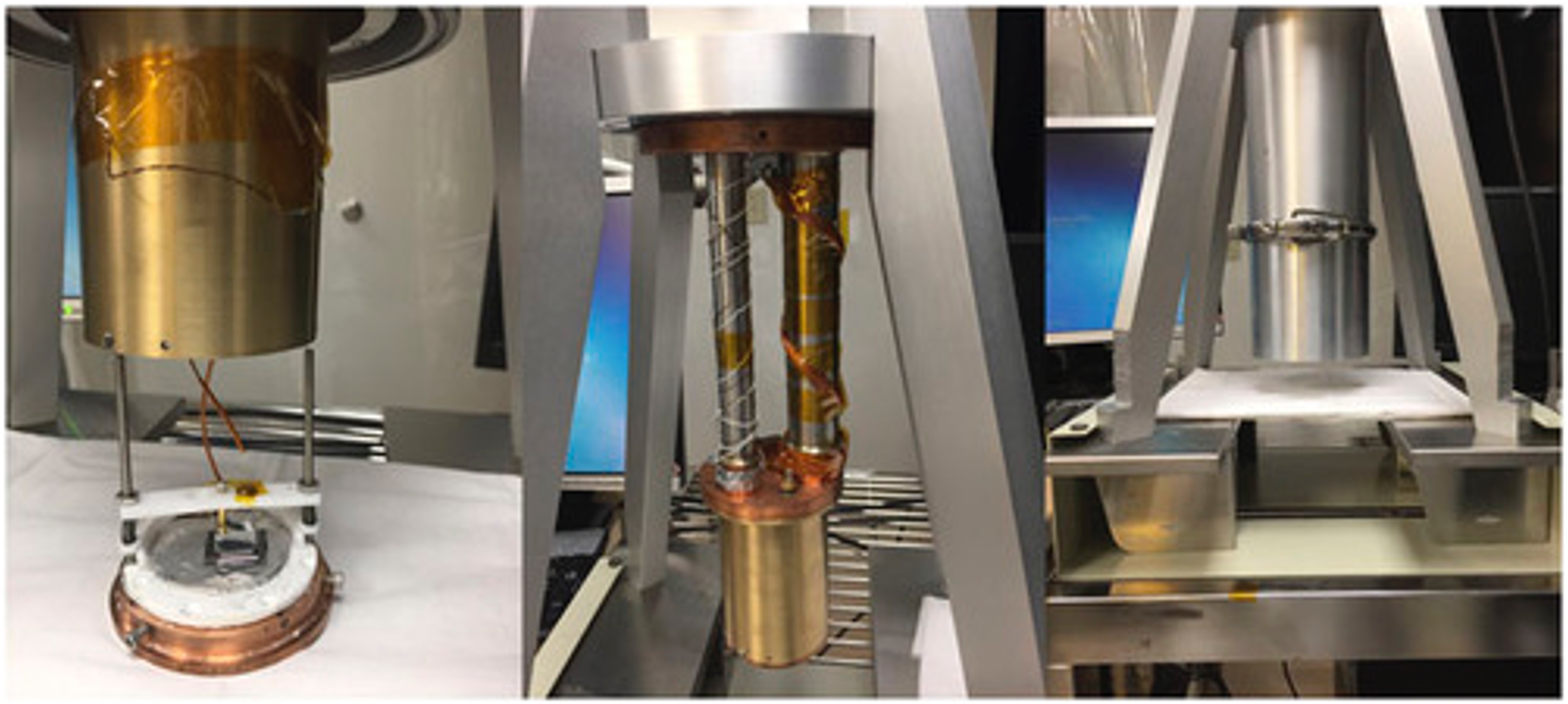}
    \caption{The detector is loaded into a pulse tube refrigerator (PTR), and two temperature sensors mounted above and below the detector are used to determine the temperature of the detector. }
    \label{fig:my_label1}
\end{figure}

To ensure optimal electrical performance, an amorphous Ge passivation layer of 600 nm was coated on the surface of the Ge crystal as the electrical contact, effectively blocking surface charges\cite{bhattarai2020experimental,wei2018investigation}. An alpha source ($^{241}Am$) was positioned near the detector inside a cryostat, and the energy deposition of $\alpha$ particles was measured. This creates localized electron-hole pairs near the top surface of the detector, and the electrons are drifted through the detector by applying a positive bias voltage to the bottom of the detector. The experimental setup for this measurement is illustrated in Figure~\ref{fig:my_label1}.

This experiment was conducted using two modes of operation. In Mode 1, the R09-02 detector was depleted at 77 K with a depletion voltage of 1200 V and an operational voltage of 1800 V. An alpha source ($^{241}Am$) emitting alpha particles with an energy of 5.3 MeV was positioned above the detector within the cryostat. The energy spectrum was measured for the energy deposition of the 5.3 MeV alpha particles, which was visible as a 3.7 MeV energy peak due to energy loss on the way to the detector's active region. This 3.7 MeV energy deposition served as a reference for the energy deposition of 5.3 MeV alpha particles in the n-type detector without charge trapping, as the detector charge trapping at 77 K with a bias of 1800 volts was negligible. The charge collection efficiency was determined by dividing the measured alpha energy peak by 3.7 MeV for a given bias voltage.

In this mode, the detector was fully depleted at a constant bias voltage of 1800 V as the temperature was decreased to 5.2 K. This allowed for the formation of electric dipole states due to space charge at 5.2 K. The data was collected with a bias voltage applied in descending order from 1800 V to 30 V at 5.2 K, with histograms of energy deposition by alpha particles recorded every 2-3 minutes for 60 minutes at each bias voltage.

In Mode 2, the detector was cooled directly to 5.2 K without any bias voltage applied. Once the temperature reached 5.2 K, a positive bias voltage was gradually applied from the bottom of the detector, causing the electrons created on the surface to be drifted across the detector under the electric field. Energy spectrum measurements were taken at different bias voltages of 30 V, 100 V, 200 V, 300 V, 450 V, 600 V, 1200 V, and 1800 V. Similar to Mode 1, data was taken for 60 minutes at each bias voltage with histograms of energy deposition by alpha particles recorded every 2-3 minutes.

\section{\label{sec:level4}Result and Discussion}
 
\begin{figure}
    \centering
    \includegraphics[width=0.5\textwidth, inner]{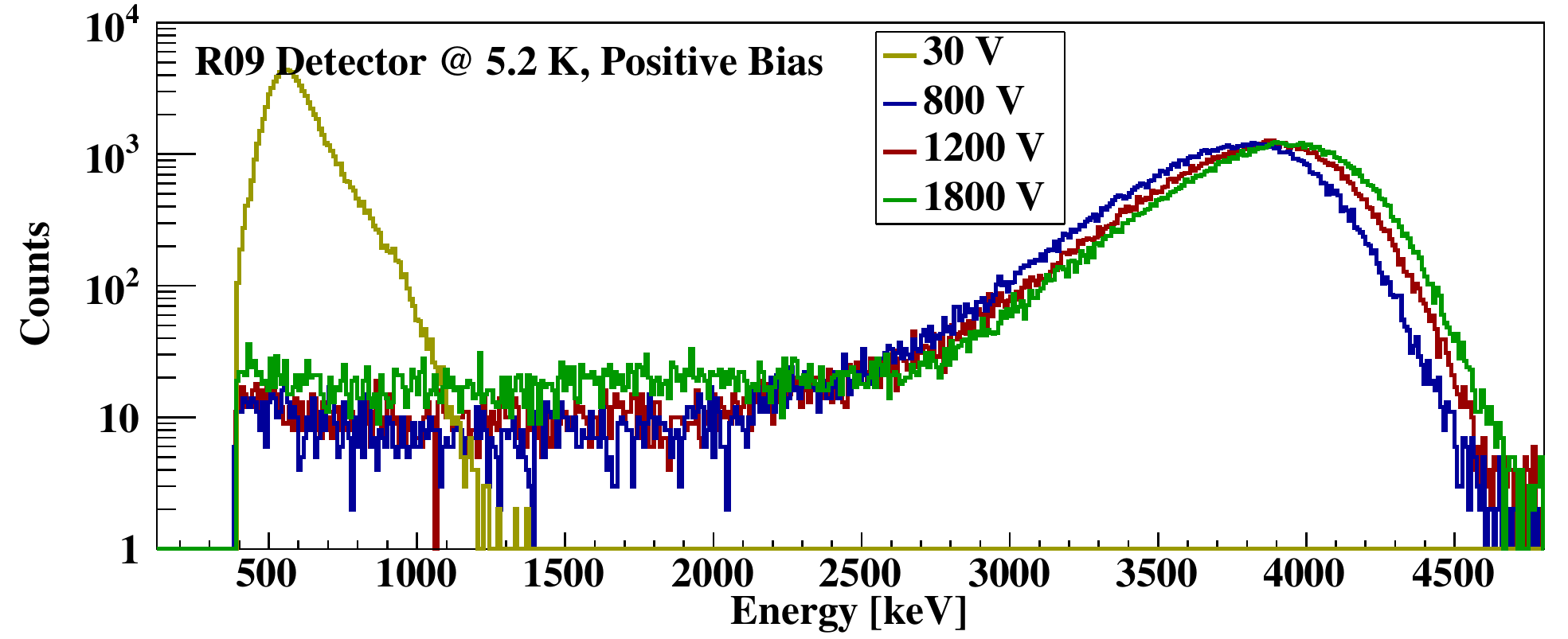}    
    \caption{The energy deposition of 5.3 MeV $\alpha$ particles in an n-type detector operating in Mode 1.} 
    \label{fig:my_label4}
\end{figure}

\begin{figure}
    \centering
    \includegraphics[width=0.5\textwidth, inner]{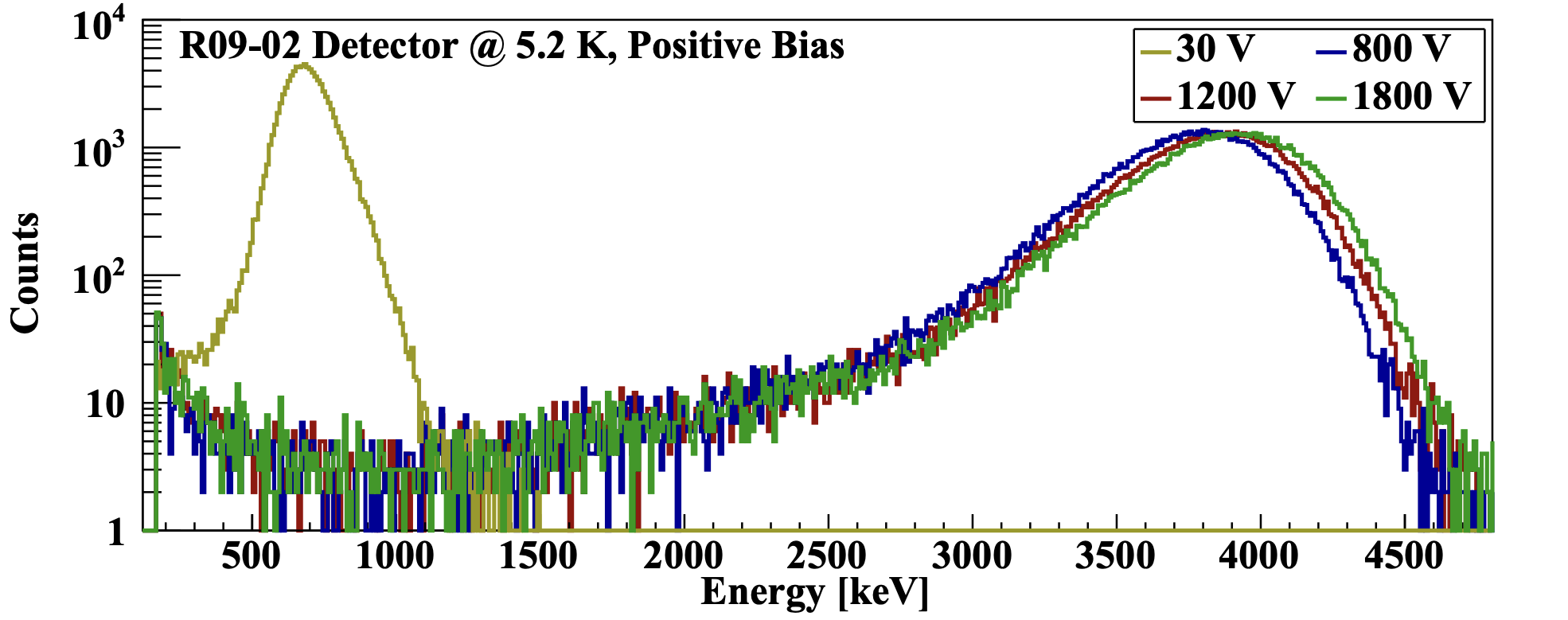}    
    \caption{The energy deposition of 5.3 MeV $\alpha$ particles in an n-type detector operating in Mode 2. }
    \label{fig:my_label3}
\end{figure}

\begin{figure}
    \centering
    \includegraphics[width=0.55\textwidth,inner]{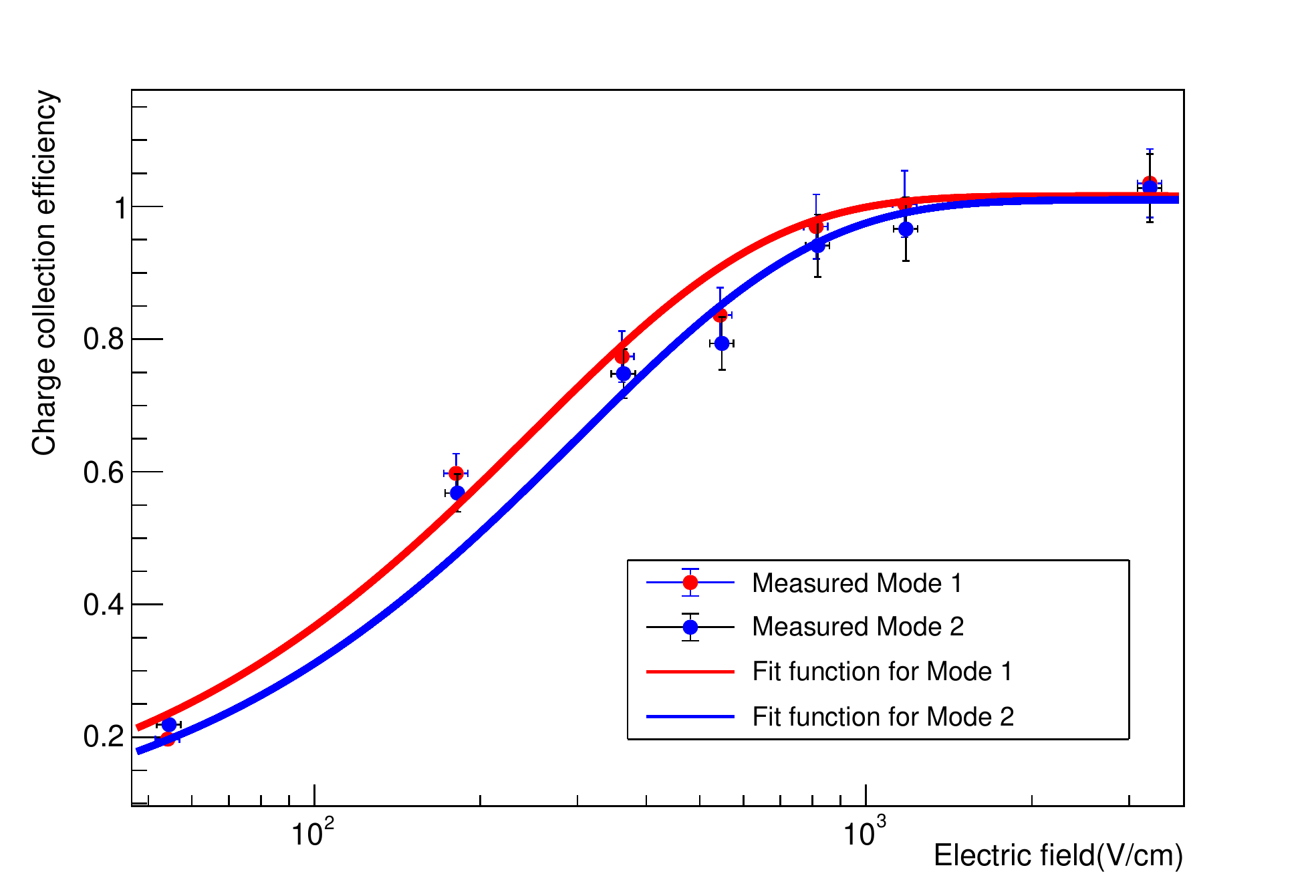}
    \caption{ The graph of charge collection efficiency ($\epsilon$) versus applied electric field ($E$) for Detector R-09 at Mode 1 and Mode 2 has been plotted, with errors taken into account. The error in $\epsilon$ is based on the measurement of the mean energy deposition, while the error in $E$ is largely influenced by the bias voltage applied. A fitting model, $\epsilon = p_0 + [(p_1 \times \text{exp}(-
    (p_2) \times E)]$, was utilized to curve-fit the data, resulting in the following fitted parameters: $p_0 = 1.01 \pm 0.008$, $p_1 = -0.973 \pm 0.001$, and $p_2 = (0.0033 \pm 0.0003) \frac{cm}{V}$ for Mode 1 and $p_0 = 1.008 \pm 0.008$, $p_1 = -0.974\pm 0.001$, and $p_2 = (0.0027 \pm 0.0003) \frac{cm}{V}$ for Mode 2 respectively.}
    \label{fig:my_label6}
\end{figure}

\begin{figure}
\includegraphics[width=0.55\textwidth, inner]{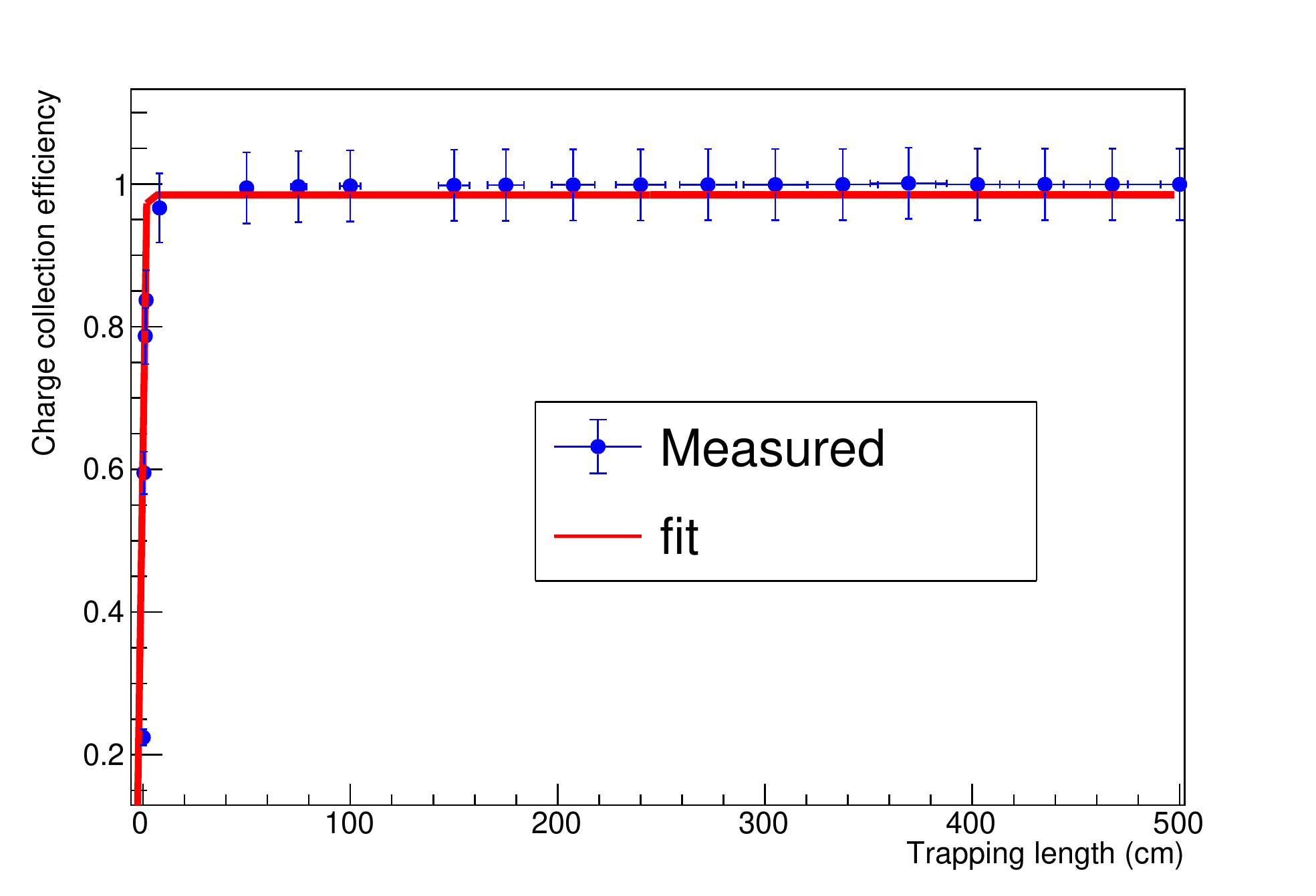}
    \caption{ The graph of charge collection efficiency ($\epsilon$) versus trapping length ($\lambda_{trap}$) for an n-type Detector R-09 has been plotted, taking into account the errors. The error in $\epsilon$ is derived from the measured mean energy deposition, while the error in $\lambda$ is calculated using the propagation of error in equation~\ref{eq:my_label7}. A fitting model, $\epsilon = \frac{p_0}{1 + (p_1 \times \text{exp}(-p_2 \times \lambda_{trap}))}$, was applied to fit the data, resulting in the following fitted parameters: $p_0 = 0.9847 \pm 0.012$, $p_1 = 4.84 \pm 0.45$, and $p_2 = (3.3\pm0.39$)/cm.}
    \label{fig:my_label2}
\end{figure}

Figures \ref{fig:my_label4} and \ref{fig:my_label3} demonstrate the energy deposition from 5.3 MeV alpha particles in the n-type detector when it operates under Mode 1 and Mode 2, respectively. The charge collection efficiency of the detector is determined by comparing the mean total energy deposited at 5.2 K with a specific bias voltage to the mean energy deposited at 77 K when the detector was depleted and operated with a bias voltage of 1800 volts. For instance, the mean energy observed at 77 K with a bias voltage of 1800 V was 3.7 MeV, while the mean energy observed at 30 V at 5.2 K was 0.725 MeV. This results in a charge collection efficiency of 19.6\% ($\epsilon$ = 0.725 MeV/3.7 MeV) in Mode 2.  Figure~\ref{fig:my_label6} shows the charge collection efficiency as a function of the applied bias voltage when the detector is operated in Mode 1 $\&$ 2. The trapping length ($\lambda_{trap}$) of the charge carriers was then calculated using equation\ref{eq:my_label7} based on the charge collection efficiencies obtained at various bias voltages and the thickness ($L$) of the detector (5.5 mm). The calculated values are presented in Figure~\ref{fig:my_label2}.

\begin{figure}
    \centering
    \includegraphics[width=0.55\textwidth, inner]{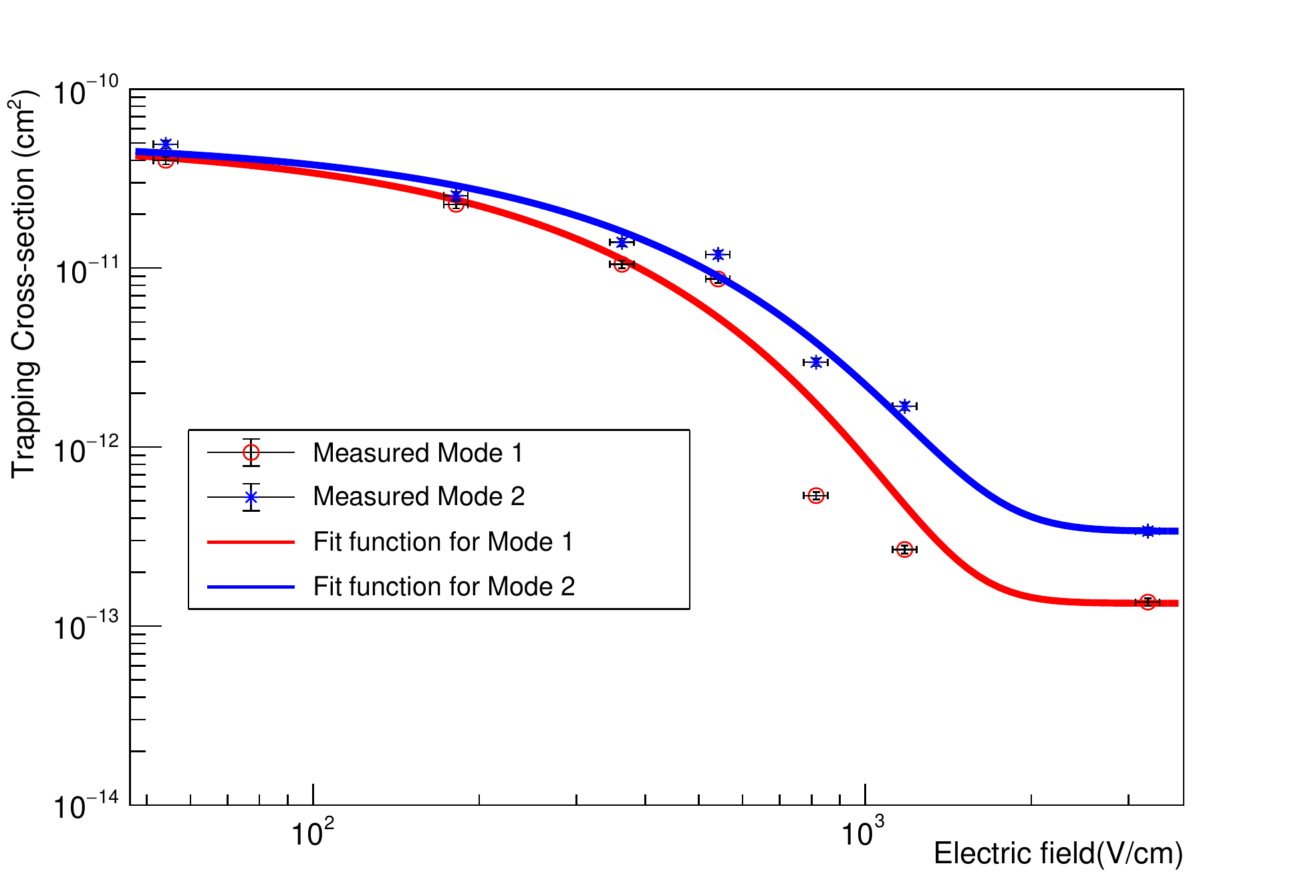}
    \caption{The graph of the variation of trapping Cross-Section ($\sigma_{trap}$) with the applied bias field ($E$) in detector R-O9 has been plotted for both Mode 1 and Mode 2, considering the errors. The error in $\sigma_{trap}$ is calculated using the propagation of error in equation~\ref{eq:my_label4} while the error associated with $E$ is primarily due to the applied bias voltage. A fitting model, $\sigma_{trap} = p_0 - [(p_1)\times \text{exp}(-p_2 \times E)]$, was used to fit the data, with the following fitted parameters for Mode 1: $p_0 = (1.34 \times 10^{-13} \pm 1.83 \times 10^{-14}$) cm$^2$, $p_1 = -(5.17 \times 10^{-11} \pm 7.4 \times 10^{-12}$) cm$^2$, and $p_2 = (0.00425 \pm 0.00014)\frac{cm}{V}$. For Mode 2, these values are: $p_0 = (3.38 \times 10^{-13} \pm 1.69 \times 10^{-14}$) cm$^2$, $p_1 = -(5.20 \times 10^{-11} \pm 5.21 \times 10^{-12}$) cm$^2$, and $p_2 = (0.000335 \pm 0.00012)\frac{cm}{V}$.}
    \label{fig:my_label5}
\end{figure}

The net impurity concentration of the detector was measured to be $7.02\times10^{10}/cm^3$ and it was operated at a temperature of 5.2 K using the two modes described earlier. These values, along with other parameters presented in equations~\ref{eq:my_label5},~\ref{eq:my_label6}, and \ref{eq:my_label7}, were utilized to calculate the trapping cross-section of the trap centers. The relationship between the trapping cross-section and the applied bias voltage is illustrated in Figure\ref{fig:my_label5}.

\begin{figure}
    \centering
    \includegraphics[width=0.55\textwidth, inner]{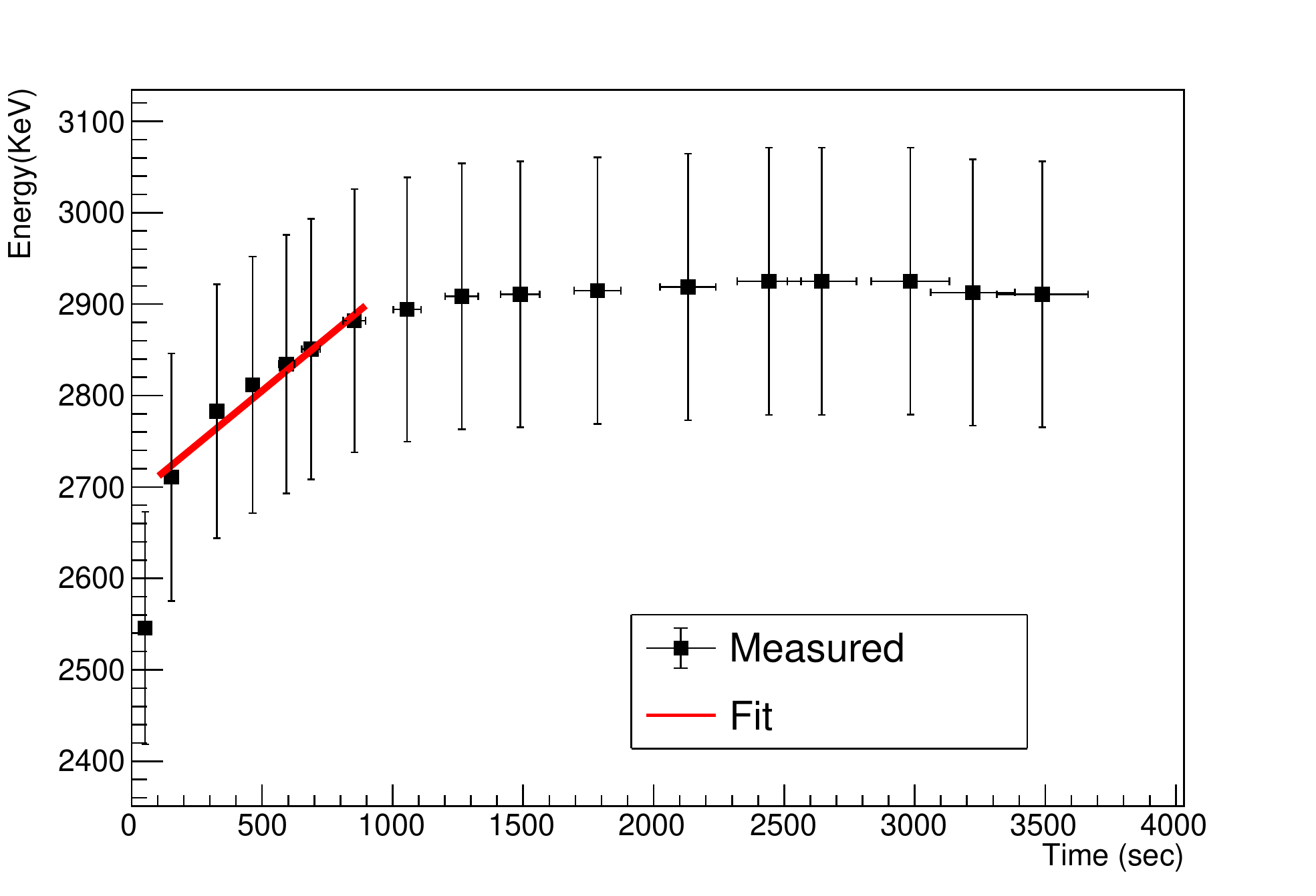}
    \caption{The mean energy deposition ($E_{dep}$) versus time ($t$) for detector R-O9 in Mode 1. As an example, the mean energy deposition ($E_{dep}$) and time ($t$) recorded for a bias voltage of 200 volts have been plotted for detector R-O9 when it is operated in Mode 1. The error in $E_{dep}$ originates from the determination of energy deposition, while the error in $t$ is primarily due to the determination of recorded time. A linear fit ($E_{dep} = p_0 \times t + p_1$) was applied to the portion of the plot where the emission of charge carriers is higher than the trapping of charge carriers. The slope ($p_0$) of the fit was calculated to be $0.235 \pm 0.025$ and the intercept ($p_1$) was $2687.09 \pm 138.8$. It is important to note that the slope represents the emission rate of charge ($e_n$) in equation~\ref{eq:my_label3}. }
    \label{fig:my_label7}
\end{figure}

To determine the charge emission rate described in equation~\ref{eq:my_label3}, we conducted a measurement of the energy deposition from $\alpha$ particles as a function of time for a given bias voltage at 5.2 K over a 60-minute interval. We recorded the histogram of the energy deposition every 2-3 minutes within this time frame. The mean value of the energy deposition was determined from the observed $\alpha$ peak. An example of this measurement is shown in Figure~\ref{fig:my_label7}, where the energy deposition versus time is plotted for a bias voltage of 200 volts.

 As demonstrated in Figure~\ref{fig:my_label7}, when the bias voltage is applied to the detector, the charge emission rate increases linearly for the first few minutes. This is due to the fact that the de-trapping through impact ionization of the dipole states or cluster dipole states outpaces the trapping of the charge carriers in the initial minutes at a given voltage. However, once the trapping and de-trapping reach a dynamic equilibrium, the energy deposition becomes constant. The slope of the portion of the plot where the emission of charge carriers is dominant provides the charge-energy emission rate per unit of time, represented as $e_n$ in equation~\ref{eq:my_label3}. By dividing $e_n$ by the binding energy of the dipole states or cluster dipole states ($E_b$), the emission rate of electrons can be obtained. These emission rates are then utilized in equation~\ref{eq:my_label3} to numerically determine the binding energy for the respective dipole states or cluster dipole states. The calculated binding energies are presented in Table~\ref{tab:my_label1}.
\begin{table*}
\begin{adjustbox}{width=1.1\textwidth,center=\textwidth}

    \begin{tabular}{|c|c|c|c|c|c|c|c|}
    \hline
    &&\multicolumn{3}{|c|}{Mode 1}&\multicolumn{3}{|c|}{Mode 2}\\
    \hline
        Bias voltage (V)& Electric field(V/cm)& Slope (eV/s)&Binding Energy(meV)&Trapping cross-section($cm^2$)&Slope(eV/s)& Binding Energy(meV)& Trapping cross  section($cm^2$)\\
        \hline
         30&$54.54\pm2.72$&$53.12\pm2.65$&$8.05\pm0.40$&$(3.99\pm0.19)\times10^{-11}$&$62.2\pm3.11$
&$8.15\pm0.40$&$(4.90\pm0.24)\times10^{-11}$\\
         \hline
         100&$181.81\pm9.09$&$236\pm11.8$&$7.09\pm0.35$&$(2.26 \pm 0.11)\times10^{-11}$&$72.7\pm3.61$&$6.58\pm0.32$&$(2.51\pm 0.13)\times10^{-11}$\\
         \hline
         200&$363.63\pm18.16$&$235.2\pm11.76$&$6.71\pm0.33$&$(1.03\pm0.05)\times10^{-11}$&$92.3\pm4.61$&$6.33\pm0.31$&$(1.37\pm0.06)\times10^{-11}$\\
         \hline
         300&$545.45\pm27.27$&$275.9\pm13.79$&$6.54\pm0.33$&$(8.59\pm0.42)\times10^{-12}$&$87.4\pm4.37$&$6.20\pm0.31$&$(1.17\pm0.06)\times10^{-11}$\\
         \hline
         450&$818.18\pm40.90$&$59.5\pm2.97$&$5.93\pm0.29$&$(5.27\pm 0.26)\times10^{-13}$&$68.2\pm3.41$&$5.47\pm0.27$&$(2.93\pm 0.14)\times10^{-12}$\\
         \hline
         650&$1181.81\pm59.05$&$29.5\pm1.47$&$5.94\pm0.28$&$(2.67\pm 0.13)\times10^{-13}$&$35.3\pm1.76$&$5.19\pm0.30$&$(1.67\pm0.08)\times10^{-12}$\\
         \hline
         1800&$3272.72\pm163.6$&$13.6\pm0.68$& $5.99\pm0.30$&$(1.35\pm0.06)\times 10^{-13}$&$19.4\pm$0.97&$4.52\pm0.22$&$(3.39\pm0.17)\times 10^{-13}$\\
         \hline
         
    \end{tabular}
    \end{adjustbox}
    \caption{The binding energy and trapping cross-section of R-09 at 5.2 K for Mode 1 and Mode 2. The errors associated with each value are either the result of measurement errors or the error calculated from the equations used in the paper.  }
    \label{tab:my_label1}
    
\end{table*}

\begin{figure}
    \centering
    \includegraphics[width=0.55\textwidth, inner]{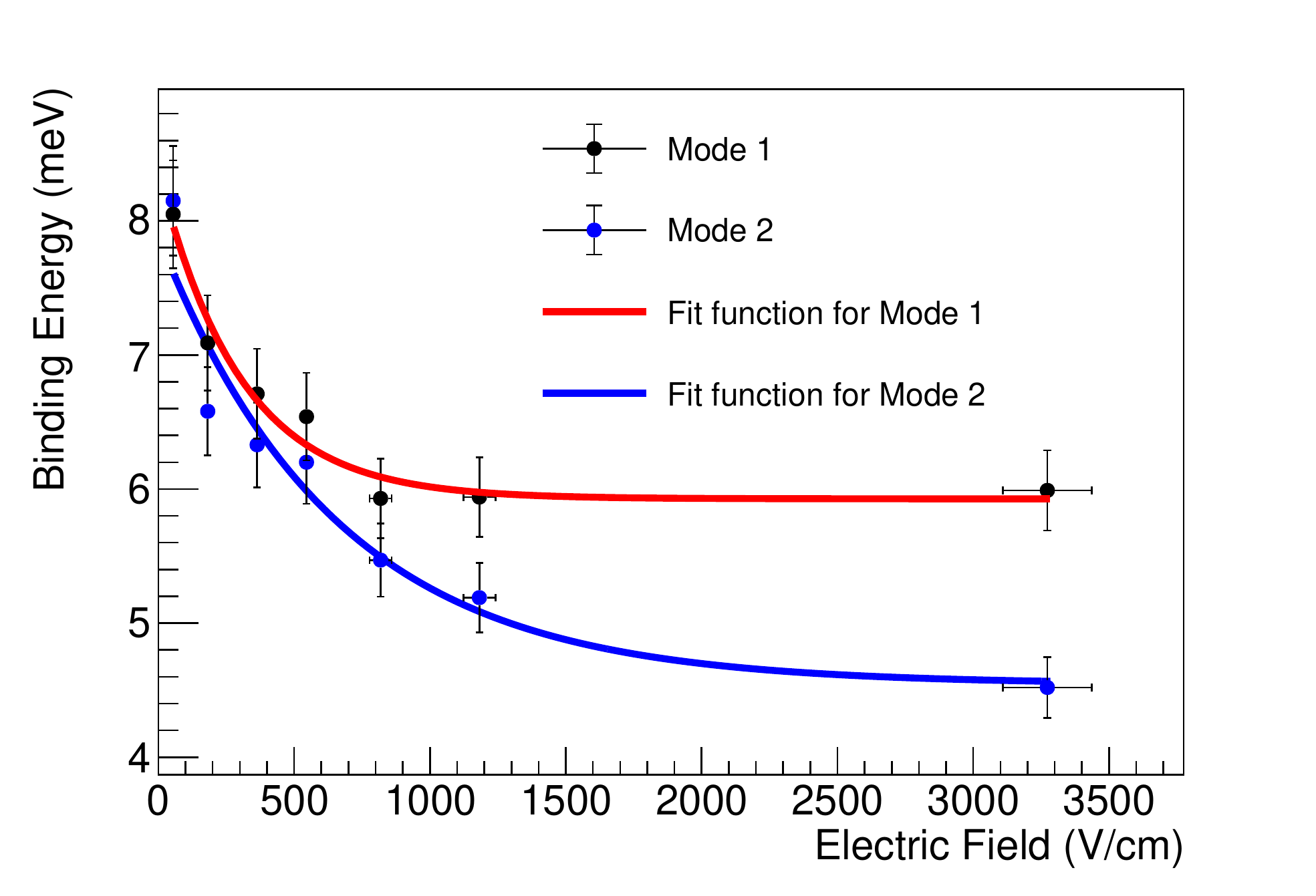}
    \caption{The binding energies of the dipole states and the cluster dipole states have been determined as a function of the applied electric field under two different operational modes, Mode 1 and Mode 2. The error in the binding energy measurement was calculated, while the error in the electric field measurement was dominated by the precision of the applied bias voltage. To analyze the data, a fit model was used, specifically $E_{B}$ =$p_0+[(p_1)\times exp(-(p_2)\times E)]$, which resulted in the following fitted parameters: For Mode 1, $p_0$ was found to be (5.927$\pm$0.219) meV, $p_1$ was (2.443$\pm$0.529) meV, and $p_2$ was (0.0033$\pm$0.001) $\frac{cm}{V}$. For Mode 2, $p_0$ was (4.545$\pm$0.248) meV, $p_1$ was (3.339$\pm$0.396) meV, and $p_2$ was (0.00154$\pm$0.0004) $\frac{cm}{V}$.}
    \label{fig:my_label8}
\end{figure}

The binding energy measured by the detector in Mode 1 pertains to the dipole states, whereas Mode 2 provides data on the binding energy of the cluster dipole states. Additionally, the binding energy values obtained at varying bias voltages demonstrate a relationship with the electric field. As shown in Figure~\ref{fig:my_label8}, the binding energies are plotted as a function of the electric field at a temperature of 5.2 K.

In Mode 1, the binding energies of the dipole states ($D^{0^{*}}$) vary from 5.99 meV to 8.05 meV depending on the electric field. When the electric field is zero, the average binding energy is calculated to be 8.369 $\pm$ 0.748 meV, which is the sum of $p_0$ + $p_1$. Similarly, the binding energies of the cluster dipole states ($D^{-^{*}}$) in Mode 2 range from 4.52 meV to 8.15 meV based on the applied electric field. At zero field, the average binding energy is 7.884 $\pm$ 0.644 meV. The results indicate that the binding energy at zero field for $D^{0^{*}}$ states is greater than that of $D^{-^{*}}$ states. Moreover, Figure\ref{fig:my_label8} reveals that $D^{-^{*}}$ states are more sensitive to the electric field than $D^{0^{*}}$ states. It should be noted that the binding energies at zero field for both $D^{0^{*}}$ states and $D^{-^{*}}$ states are lower than the binding energies of ground state impurity atoms in a Ge detector, which typically fall within the range of 10 meV.

\section{\label{sec:level3}Conclusions}
Our study of binding energies and trapping cross-sections in an n-type Ge detector operating at a low temperature has revealed valuable insights. Our measurements indicate that the binding energy of dipole states is 8.369 $\pm$ 0.748 meV and the binding energy of cluster dipoles is 7.884 $\pm$ 0.644 meV, both of which are lower than the typical binding energy (around 10 meV) of ground state impurities in Ge. We found that at a temperature of 5.2 K, the thermal energy of 0.448 meV is much lower than these binding energies, indicating that the corresponding cluster dipole states and dipole states are thermally stable at a temperature of 5.2 K. The application of an electric field causes the smaller binding energy of cluster dipoles to result in increased de-trapping via impact ionization when compared to dipole states. The trapping cross section, which ranges from $3.99\times10^{-11} cm^2$ to $1.35\times10^{-13} cm^2$, is primarily influenced by the electric field. Our findings further demonstrate that the binding energy and trapping cross-section decrease as the electric field within the detector increases. These low binding energies suggest the potential for developing a low-threshold detector using appropriately doped impurities in Ge for low-mass dark matter searches.

\section{Acknowledgments}
The authors would like to thank Mark Amman for his instructions on fabricating planar detectors. We would also like to thank the Nuclear Science Division at Lawrence Berkeley National Laboratory for providing us with a testing cryostat. This work was supported in part by NSF OISE 1743790, DE-SC0004768, and a governor's research center supported by the State of South Dakota.

\bibliography{sanjay.bib}

\end{document}